\documentclass[twocolumn,showkeys,aps,prb,showpacs]{revtex4-1}
\usepackage{graphicx}
\usepackage[CJKbookmarks,dvipdfm,colorlinks,linkcolor=blue,citecolor=blue]{hyperref}

\begin{document}

\title{ Predicted Janus SnSSe monolayer: a comprehensive first-principle study}

\author{San-Dong Guo and Xiao-Shu Guo}
\affiliation{School of Electronic Engineering, Xi'an University of Posts and Telecommunications, Xi'an 710121, China}
\begin{abstract}
The Janus structure, by combining  properties of different transition metal dichalcogenide (TMD) monolayers
in a single polar material, has attracted increasing research interest because of  their
particular structure and  potential application in electronics, optoelectronics and piezoelectronics.
In this work, Janus SnSSe monolayer is predicted by means of first-principles calculations, which exhibits  dynamic and mechanical stability.
By using generalized gradient approximation (GGA) plus spin-orbit coupling (SOC), the Janus SnSSe monolayer  is found to be an indirect band-gap semiconductor, whose gap  can easily be tuned by strain. High  carrier
mobilities are obtained for SnSSe monolayer, and the hole mobility is higher than the electron mobility.
For SnSSe monolayer, a uniaxial strain in the basal plane can induce both strong in-plane and much weaker out-of-plane piezoelectric polarizations, which reveals the potential as a piezoelectric two-dimensional (2D) material. The high absorption coefficients
in the visible light region are observed, suggesting a potential  photocatalytic application.  Calculated results show that SnSSe monolayer has very high power factor, making it a promising candidate for thermoelectric applications.
Our works reveal that the Janus SnSSe structure can
be fabricated with unique electronic, optical,  piezoelectric and transport properties, and can motivate related experimental works.

\end{abstract}

\pacs{71.20.-b, 77.65.-j, 72.15.Jf, 78.67.-n ~~~~~~~~~~~~~~~~~~~~~~~~~~~~~~~~~~~Email:sandongyuwang@163.com}

\maketitle

\section{Introduction}
After the exploration of graphene\cite{q6}, the search for new 2D materials with unique structures and electronic properties have been in progress due to potential  applications in the field of optoelectronics and energy conversion and storage.
Many  2D materials have been synthesized experimentally or predicted theoretically, such as TMD, group-VA, group IV-VI and group-IV  monolayers\cite{q7,q8,q9,p1,q6-1,q6-2,q10,q11}. Among these, TMD monolayers  are prominent due to their finite direct band
gaps, large on-off ratios, high carrier mobilities and   strong optoelectronic responses\cite{q6-3,q6-4}.
Due to the lack of inversion symmetry, 2H TMD monolayers posses  a spin-orbit splitting in the valence band,  allowing for valley-selective excitation
of carriers\cite{q6-5}. Using symmetry analysis, the  piezoelectric coefficients of TMD monolayers have been calculated  with density functional
perturbation theory (DFPT)\cite{q6-6}.
Many studies of  heat transport properties of  TMD monolayers  have been reported, and strain effects on their  heat transport properties  have been investigated\cite{q12,q13,q20,q22,q23,l11}.

\begin{figure}
  \includegraphics[width=7.0cm]{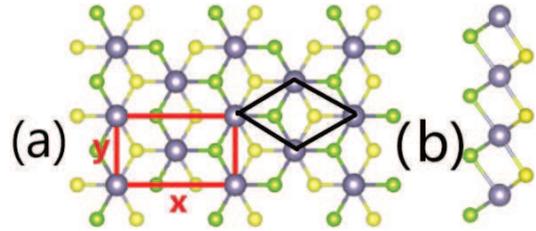}
  \caption{(Color online) The top view (a) and side view (b) crystal structure of Janus SnSSe monolayer. The large balls represent Sn atoms, and the  small balls for S/Se atoms. The black and red lines represent the rhombus primitive cell  and the rectangle supercell.}\label{t0}
\end{figure}

In general, the two-layer chalcogen atoms in TMD monolayers  are same element. If the two-layer chalcogen atoms are different, Janus  monolayer will be constructed, and then the symmetry will be reduced.
Recently, Janus monolayer MoSSe has been successfully synthesised  by different
experimental strategies, leading to an out-of-plane structural asymmetry\cite{p1,p1-new}.
Compared to commonly used 2D materials, the out-of-plane piezoelectric effects of monolayer and multilayer Janus TMD  MXY (M = Mo, W; X/Y = S, Se, Te) have been achieved, and out-of-plane piezoelectric
coefficient  is found to be strongest in multilayer MoSTe \cite{p2,p2-new}.
The carrier mobility of monolayer MoSSe can be improved by building bilayer or trilayer structures \cite{p2-1}.
The mechanical and electronic properties of Janus monolayer  MXY (M=Ti, Zr, Hf, V, Nb, Ta, Cr, Mo, W; X/Y=S, Se, Te) have been systematically investigated using density functional theory (DFT)\cite{p2-1-1}. A significant Rashba spin splitting can be induced by intrinsic out-of-plane built-in electric field in monolayer Janus WSeTe\cite{p2-2}. Janus MoSSe and PtSSe  monolayers are predicted to be   potential wide solar-spectrum water-splitting photocatalyst\cite{p1-1,p1-2}.
The electronic and optical properties of  MoSSe-WSSe vertical and lateral heterostructures  have been studied  by the first-principles calculations\cite{p3}. The lattice thermal conductivities of MoSSe/ZrSSe/PtSSe monolayer are predicted by linearized phonon Boltzmann equation, which are much lower than that of the  $\mathrm{MoS_2}$/$\mathrm{ZrS_2}$/ $\mathrm{PtS_2}$ monolayers\cite{p4,p4-1,p4-2}.
\begin{figure}
  \includegraphics[width=8cm]{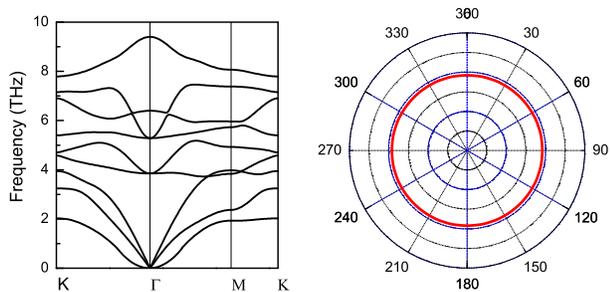}
\caption{(Color online) The phonon band dispersion  of  SnSSe monolayer, and the orientation angle-dependent 2D Young's modulus $C_{2D}(\theta)$. }\label{t1}
\end{figure}

In this work, inspiring from the already synthesized  1T structure of $\mathrm{SnS_2}$/$\mathrm{SnSe_2}$\cite{q6-1,q6-2}, we systematically investigate the electronic structures, carrier mobilities, piezoelectric properties, optical properties and  transport coefficients   of Janus SnSSe monolayer.
We predict that dynamically and mechanically  stable SnSSe monolayer has
distinctive electronic, optical,  piezoelectric and transport properties.

\begin{table}
\centering \caption{For SnSSe monolayer, the lattice constants $a_0$ ($\mathrm{{\AA}}$), the elastic constants $C_{ij}$, shear modulus
$G_{2D}$,  Young's modulus $C_{2D}$ in $\mathrm{Nm^{-1}}$,  Poisson's ratio $\nu$
dimensionless and the gaps with GGA and GGA+SOC (eV). }\label{tab0}
  \begin{tabular*}{0.48\textwidth}{@{\extracolsep{\fill}}cccc}
  \hline\hline
$a_0$& $C_{11}=C_{22}$ &  $C_{12}$& $G_{2D}$\\\hline
3.78 &62.50&17.68&22.41\\\hline\hline
$C_{2D}$& $\nu$& Gap& Gap-SOC\\\hline
57.50&0.28&0.95 &0.83\\\hline\hline
\end{tabular*}
\end{table}

\section{Computational detail}
Within the DFT \cite{1}, a full-potential linearized augmented-plane-waves method
is used to investigate electronic structures  of  SnSSe monolayer by using  WIEN2k  code\cite{2}. We use  popular GGA of Perdew, Burke and  Ernzerhof  (GGA-PBE)\cite{pbe} as the exchange-correlation potential, and the SOC was included self-consistently \cite{10,11,12,so}.
 To attain reliable results, we use  a 30 $\times$ 30 $\times$ 1 k-point meshes in the first Brillouin zone (BZ) for the self-consistent calculation,  make harmonic expansion up to $\mathrm{l_{max} =10}$ in each of the atomic spheres, and set $\mathrm{R_{mt}*k_{max} = 8}$. The charge  convergence threshold is set as  $0.0001|e|$ per formula unit, in which  $e$ is the electron charge.  The electronic transport coefficients of Janus SnSSe monolayer  are performed through solving Boltzmann
transport equations within the constant scattering time approximation (CSTA), as implemented in
the BoltzTrap package\cite{b}. To attain  accurate transport coefficients, a 110 $\times$ 110 $\times$ 1 k-point meshes is used in the first BZ for the energy band calculation. The phonon calculations are
carried out by VASP+PHONOPY code\cite{pv1,pv2,pv3,pv5} with a supercell
of 6$\times$6$\times$1 using the finite displacement method. The kinetic energy cutoff is
set to 500 eV.  We calculate the elastic tensor
 and   piezoelectric stress tensors within DFPT\cite{pv6}, using the VASP code.

Due to 3D periodic boundary conditions, the 2D elastic coefficients $C^{2D}_{ij}$
 and   piezoelectric stress coefficients $e^{2D}_{ij}$
have be renormalized by the the length of unit cell along z direction ($Lz$):  $C^{2D}_{ij}$=$Lz$$C^{3D}_{ij}$ and $e^{2D}_{ij}$=$Lz$$e^{3D}_{ij}$.
For 2D materials, it is noted that the calculated optical properties and  electrical conductivity  also depend on $Lz$\cite{pv7,2dl}, which  should be normalized by multiplying $Lz/d$,  where $d$ for the thickness of 2D material. However, the $d$  is not well defined like graphene.  In this work, the $Lz$=20 $\mathrm{{\AA}}$  is used as $d$.

\begin{figure}
  \includegraphics[width=8cm]{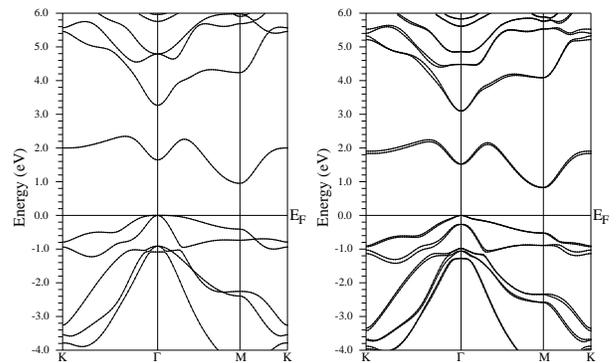}
\caption{The energy band structures  of  SnSSe monolayer  using GGA (Left) and GGA+SOC (Right). }\label{t1-1}
\end{figure}

\section{Structure and stability}
The crystal structure of Janus SnSSe monolayer is shown  in \autoref{t0}, and the rhombus primitive cell  and the rectangle supercell are marked.
It is useful for the calculations of carrier mobility and piezoelectric coefficients to use rectangle supercell.
With Pt sandwiched between the S and Se layers,   the symmetry of Janus SnSSe
monolayer (No.156) is lower than that of the $\mathrm{SnS_2}$ or $\mathrm{SnSe_2}$  monolayer (No.164) due to
the lack of  the reflection symmetry with respect to the central  Sn atoms.
To avoid spurious interaction between neighboring periodic images, the
vacuum space along the z direction is set to more than 18 $\mathrm{{\AA}}$ to separate
periodic images. The optimized lattice constants of Janus SnSSe monolayer is 3.784 $\mathrm{{\AA}}$ with GGA, being between the ones of $\mathrm{SnS_2}$ (3.69  $\mathrm{{\AA}}$) and $\mathrm{SnSe_2}$ (3.86 $\mathrm{{\AA}}$)\cite{la1}.
To confirm the stability of SnSSe monolayer, the  phonon dispersion of
Janus  monolayer SnSSe is plotted in \autoref{t1}, which shows no
imaginary modes, suggesting that monolayer SnSSe is dynamically stable.
Due to three atoms in the unit cell, the 3 acoustic and 6 optical phonon branches are observed.
Due to  $D_{3h}$ symmetry of SnSSe  monolayer,  the optical
lattice-vibration modes at $\Gamma$ point can be  defined as:
\begin{equation}\label{e1}
\mathrm{\Gamma_{optical}\equiv A^{"}_2(IR)+A^{'}_1(R)+E^{'}(IR+R)+E^{"}(R)}
\end{equation}
where  IR and R mean  infrared- and Raman-active mode, respectively.
The optical phonon frequencies of SnSSe  at $\Gamma$ point are 9.40 THz ($\mathrm{A^{"}_2}$), 6.40 THz ($\mathrm{A^{'}_1}$), 5.29 THz ($\mathrm{E^{'}}$) and 3.85 THz ($\mathrm{E^{"}}$).
The mechanical
stability of SnSSe monolayer can be examined by elastic
constants $C_{ij}$.
Due to hexagonal symmetry,  two  independent elastic
constants  $C_{11}$ and $C_{12}$ are calculated, the  $C_{11}$=$C_{22}$=62.50 $\mathrm{Nm^{-1}}$ and $C_{12}$=17.68 $\mathrm{Nm^{-1}}$, and the $C_{66}$=($C_{11}$-$C_{12}$)/2=22.41 $\mathrm{Nm^{-1}}$,  which satisfy the  Born  criteria of mechanical stability\cite{ela}, namely
\begin{equation}\label{e1}
C_{11}>0,~~ C_{66}>0
\end{equation}
The Young's modulus $C_{2D}(\theta)$ can be calculated  on the basis of the elastic constants. The
Young's modulus $C_{2D}(\theta)$ along the in-plane  $\theta$ can be expressed as follows\cite{ela1}:
\begin{equation}\label{c2d}
C_{2D}(\theta)=\frac{C_{11}C_{22}-C_{12}^2}{C_{11}sin^4\theta+Asin^2\theta cos^2\theta+C_{22}cos^4\theta}
\end{equation}
where $A=(C_{11}C_{22}-C_{12}^2)/C_{66}-2C_{12}$. The  $C_{2D}(\theta)$ is shown in \autoref{t1}, which shows strong mechanical isotropy for SnSSe monolayer. The calculated $C_{2D}(\theta)$ is 57.50 $\mathrm{Nm^{-1}}$, which is smaller than one of MoSSe (113.0 $\mathrm{Nm^{-1}}$)\cite{p4}, indicating that monolayer SnSSe is less rigid than MoSSe monolayer.
The Poisson's ratio $\nu(\theta)$ is also isotropic, and the calculated  $\nu$ is 0.28.  The shear modulus $G_{2D}$ equals to  $C_{66}$. The related data are summarized in \autoref{tab0}.
The phonon calculations and
elastic constants prove that Janus monolayer SnSSe is dynamically and mechanically stable. Monolayer $\mathrm{SnX_2 (X = S, Se)}$ have been
synthesized\cite{q6-1,q6-2}, and MoSSe monolayer have also been successfully fabricated\cite{p1,p1-new}. A similar growth method to
that of monolayer MoSSe can be used to realize Janus monolayer SnSSe.

\section{Electronic structure}
It has been proved that the SOC has  important effects on electronic structures of Janus TMD monolayers\cite{q5}. Thus, the SOC is included  for electronic structure  calculations of Janus SnSSe monolayer, and the related energy bands  with GGA and GGA+SOC are plotted in \autoref{t1-1}.
 By using GGA, an indirect gap of 0.95 eV with valence band maximum (VBM) at the $\Gamma$ point  and conduction band minimum (CBM) at the M point.
It is noted that the bands   around  VBM  are very flat, which can produce very large effective masses of VBM.
When including SOC, the gap reduces to 0.83 eV, and a  spin-orbit splitting  at  the $\Gamma$ point in the valence bands near the Fermi level
is 0.266 eV.

\begin{figure}
   \includegraphics[width=7.0cm]{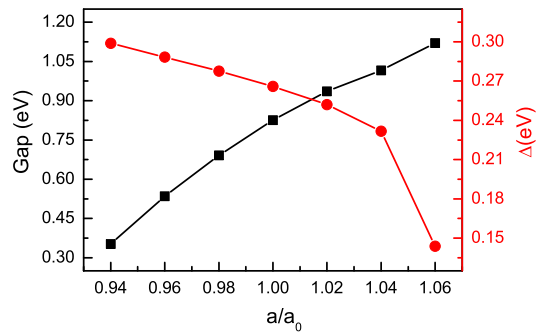}
  \caption{(Color online) The energy band gap (Gap) and spin-orbit splitting value ($\Delta$) at high symmetry $\Gamma$ point as a function of  $a/a_0$ by using GGA+SOC.}\label{t2}
\end{figure}

\begin{figure*}
  \includegraphics[width=15cm]{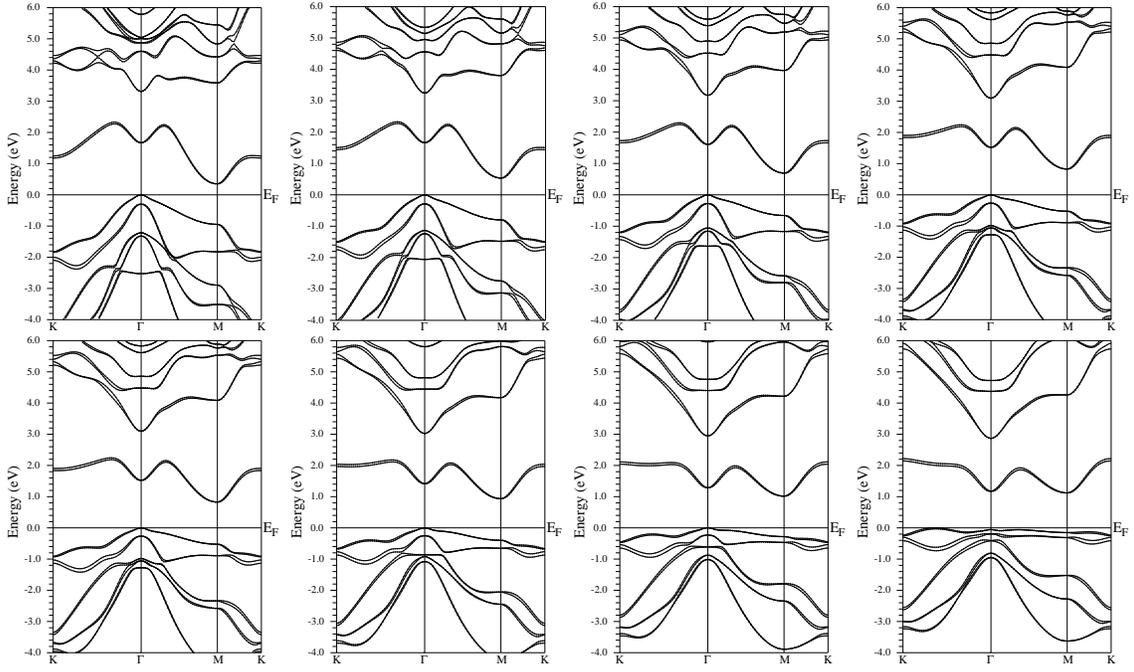}
\caption{With $a/a_0$ changing from 0.94 to 1.00 (Top) and 1.00 to 1.06 (Bottom), the energy band structures  of SnSSe monolayer using GGA+SOC, and the interval is 0.02. }\label{t3}
\end{figure*}

 Strain can effectively tune the electronic structures  of  Janus TMD  monolayers\cite{q5}. Here,  the effects of  biaxial strain on the electronic structures of   SnSSe monolayer are  examined. To simulate biaxial strain, $a/a_0$ is used with $a/a_0$$<$1 ($a/a_0$$>$1) being compressive (tensile) strain, where $a$ and $a_0$ are the strained and  unstrained lattice constant, respectively.
The energy band gap (Gap)  and   spin-orbit splitting value at  $\Gamma$ point ($\Delta$) as a function of $a/a_0$   are shown in \autoref{t2}, and the related energy band structures with $a/a_0$ from 0.94 to 1.06 are  plotted  in \autoref{t3}.  It is found that  the energy band gap   increases monotonously with $a/a_0$ changing   from 0.94 to 1.06, which is different from one of other Janus TMD  monolayers (up-and-down trend)\cite{q5}.
 With $a/a_0$  from 0.94  to 1.04, the $\Delta$ decrease monotonously, and then a rapid decrease from 1.04 to 1.06.
 It is because that the VBM changes from $\Gamma$ point to another point along $\Gamma$-K direction.
 The  strain can tune the positions of conduction band extrema (CBE), and the second CBE changes from K point to $\Gamma$ point, when strain varies from compressive strain to tensile one.
 The conduction bands  convergence can be realized by tensile strain, which can be observed at $a/a_0$=1.06 point.
 The more localized valence bands can be induced by tensile strain. These can  produce important influence on electronic transport properties of SnSSe monolayer.

\section{Carrier mobility}
As is well known, the carrier mobility of a semiconductors is  an important factor for   great potential in the electronics and optoelectronics.
To evaluate the carrier mobility, the deformation potential (DP) theory has been proposed by Bardeen and Shockley\cite{dp}, which can be used to calculate the intrinsic carrier mobility of 2D materials. In this theory, the carrier
mobility of a 2D material  ($\mu_{2D}$) is defined as:
\begin{equation}\label{u2d}
  \mu_{2D}=\frac{e\hbar^3C_{2D}}{K_BTm^*m_dE_l^2}
\end{equation}
where $T$ is temperature, and  $m^*$ is the effective mass in the transport direction, and $m_d=\sqrt{m_xm_y}$ is the average effective mass.
The $C_{2D}$ is the elastic modulus  derived from $C_{2D}$=$[\partial^2E/\partial^2\delta]/S_0$, where
$E$ is the total energy after applying uniaxial strain ($\delta=\Delta l/l_0$) and
$S_0$ is the area at equilibrium. In addition, $E_l$ is the DP constant defined by $E_l=\Delta E/\delta$
, where $\Delta E$ is the energy shift of the band edge of CBM or VBM with respect
to the vacuum level. After attaining $\mu_{2D}$,  the  relaxation
time $\tau$ can be calculated by:
\begin{equation}\label{t}
    \tau=\mu_{2D}m^*/e
\end{equation}

\begin{table*}
\centering \caption{For SnSSe monolayer, elastic modulus ($C_{2D}$), effective mass ($m^*$) using GGA+SOC  with  GGA results given in parentheses, deformation potential ($E_l$), carrier mobility ($\mu_{2D}$)   and relaxation
time ($\tau$) at 300 K.}\label{tab3}
  \begin{tabular*}{0.96\textwidth}{@{\extracolsep{\fill}}ccccccc}
  \hline\hline
Carrier type&    &$C_{2D}$ ($\mathrm{Nm^{-1}}$) & $m^*$ & $E_l$ (eV)& $\mu_{2D}$ ($\mathrm{cm^2V^{-1}s^{-1}}$)&$\tau$ (s)\\\hline\hline
Electrons   & x&  57.50&        0.74 (0.73)&       3.94         & 258.63&  $1.08\times10^{-13}$    \\
            &y&    57.50&       0.23 (0.24) &      -7.49          & 224.71   & $2.99\times10^{-14}$                                             \\
Holes   & x&       57.50&      -0.36 (-31.44)&     -4.01          &580.20     & $1.19\times10^{-13}$                                            \\
             &y&   57.50&      -0.37 (-12.79)&     -4.10          &548.74      & $1.14\times10^{-13}$                             \\\hline\hline
\end{tabular*}
\end{table*}
\begin{figure*}
  \includegraphics[width=12cm]{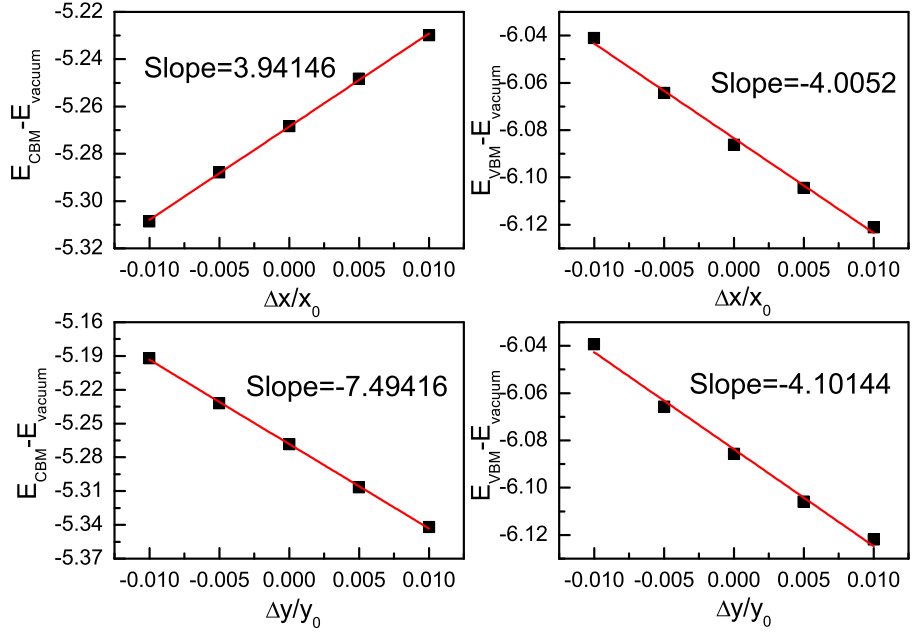}
  \caption{(Color online)Using GGA+SOC, the band energies of the VBM and CBM of SnSSe monolayer with respect to the vacuum energy as a function of lattice dilation along both x and y directions. The red solid lines are linear fitting curves. The fitted slopes are shown, which corresponds to the DP.}\label{t4}
\end{figure*}

By using   DP theory,  the carrier mobilities of SnSSe monolayer in both zigzag and armchair
directions are calculated, and the rectangular supercell is used with armchair and zigzag being defined as x and y directions in \autoref{t0}.
The calculated effective masses for electrons  and holes with GGA and GGA+SOC are shown in \autoref{tab3}. It is found that  the effective masses for electrons between GGA and GGA+SOC are very close, but the ones for holes are very different. These can be explained from their energy structures in \autoref{t1-1}. When including SOC, the bands near CBM  has little change with respect to GGA results, but the flat bands around VBM  disappear.
Thus, it is very important for calculating effective masses of holes   to consider SOC.
The DP constant $E_l$ is calculated by linearly fitting the band energies of the VBM and CBM  with respect to the vacuum energy as a function of $\Delta x/x$ and $\Delta y/y$, which is plotted in \autoref{t4}. On the basis of the calculated effective mass, elastic
constant, and deformation potential constant, the carrier mobility and relaxation
time  for the electrons and holes of monolayer SnSSe along x and y directons are calculated, which are shown \autoref{tab3}.
Predicted carrier mobilities between  electrons and holes show very strong anisotropy, and the weakly anisotropic ones along x and y directions for the same carrier type are observed.
The hole carrier mobilities of SnSSe monolayer are higher than those of monolayer $\mathrm{SnS_2}$ (187.44 $\mathrm{cm^2V^{-1}s^{-1}}$)  and  $\mathrm{SnSe_2}$ (115.65 $\mathrm{cm^2V^{-1}s^{-1}}$)\cite{la1}, but its electron carrier mobilities are lower than their ones.
The calculated $\tau$ is useful for attaining the electronic  transport coefficients of SnSSe monolayer.

\section{Piezoelectric properties}
In noncentrosymmetric materials, the piezoelectric effect can be  produced, when  mechanical stress is applied. In 2D materials,
 the third-rank piezoelectric stress tensors  $e_{ijk}$ and strain tensor $d_{ijk}$, employing Voigt notation, can be connected by elastic   tensor\cite{q6-6}:

 \begin{equation}\label{pe0}
    e=dC
 \end{equation}

 \begin{equation}\label{pe}
 e=
  \left(
    \begin{array}{ccc}
      e_{11} & e_{12} & e_{16} \\
     e_{21} & e_{22} & e_{26} \\
      e_{31} & e_{32} & e_{36} \\
    \end{array}
  \right)
   \end{equation}

  \begin{equation}\label{pe}
  d=
   \left(
    \begin{array}{ccc}
      d_{11} & d_{12} & d_{16} \\
      d_{21} & d_{22} & d_{26} \\
      d_{31} & d_{32} & d_{36} \\
    \end{array}
  \right)
   \end{equation}

  \begin{equation}\label{pe}
  C=
   \left(
    \begin{array}{ccc}
      C_{11} & C_{12} & C_{16} \\
     C_{21} & C_{22} &C_{26} \\
      C_{61} & C_{62} & C_{66} \\
    \end{array}
  \right)
 \end{equation}
Performing symmetry analysis,  the number of independent tensor coefficients can be further reduced. Due to a $3m$ point-group symmetry of SnSSe
 monolayer, the  piezoelectric stress   and strain tensors, and elastic   tensor become:
 \begin{equation}\label{pe1-1}
 e=\left(
    \begin{array}{ccc}
      e_{11} & -e_{11} & 0 \\
     0 & 0 & -e_{11} \\
      e_{31} & e_{31} & 0 \\
    \end{array}
  \right)
    \end{equation}

  \begin{equation}\label{pe1-2}
  d= \left(
    \begin{array}{ccc}
      d_{11} & -d_{11} & 0 \\
      0 & 0 & -2d_{11} \\
      d_{31} & d_{31} &0 \\
    \end{array}
  \right)
\end{equation}

 \begin{equation}\label{pe1-3}
   C=\left(
    \begin{array}{ccc}
      C_{11} & C_{12} & 0 \\
     C_{12} & C_{11} &0 \\
      0 & 0 & (C_{11}-C_{12})/2 \\
    \end{array}
  \right)
\end{equation}
\begin{table}
\centering \caption{Piezoelectric coefficients $e_{11}(d_{11})$ and $e_{31} (d_{31})$ of SnSSe and MoSSe Monolayers, and the unit is $10^{-10}$C/m (pm/V). }\label{tab5}
  \begin{tabular*}{0.48\textwidth}{@{\extracolsep{\fill}}ccccc}
  \hline\hline
MXY & $e_{11}$ & $d_{11}$& $e_{31}$&$d_{31}$\\\hline\hline
SnSSe&1.009& 2.251&-0.091&-0.114\\\hline
MoSSe& 3.890&      4.238   &0.418 & 0.285                                \\\hline\hline
\end{tabular*}
\end{table}

Here, $d_{11}$ and $d_{31}$ are derived by  \autoref{pe0}, \autoref{pe1-1}, \autoref{pe1-2} and \autoref{pe1-3}:
\begin{equation}\label{pe2}
    d_{11}=\frac{e_{11}}{C_{11}-C_{12}}~~~and~~~d_{31}=\frac{e_{31}}{C_{11}+C_{12}}
\end{equation}

The nonprimitive orthorhombic unit cells of SnSSe monolayer (in \autoref{t0}) is used  as the
computational unit cell for $e_{ij}$ .  The calculated $e_{ij}$ and $d_{ij}$ of SnSSe monolayer are shown
in \autoref{tab5}, along with ones of already synthesized MoSSe monolayer. Calculated results show that the piezoelectric coefficients of
SnSSe are lower than ones of MoSSe. However, the $d_{11}$ is comparable  with one of $\alpha$-quartz ($d_{11}$=2.27 pm/V)\cite{yd3} commonly used
3D piezoelectric material. It is noted  that $\mathrm{SnS_2}$ and $\mathrm{SnSe_2}$ monolayers  display inversion symmetry and therefore no
piezoelectricity.

\begin{figure}
    \includegraphics[width=7cm]{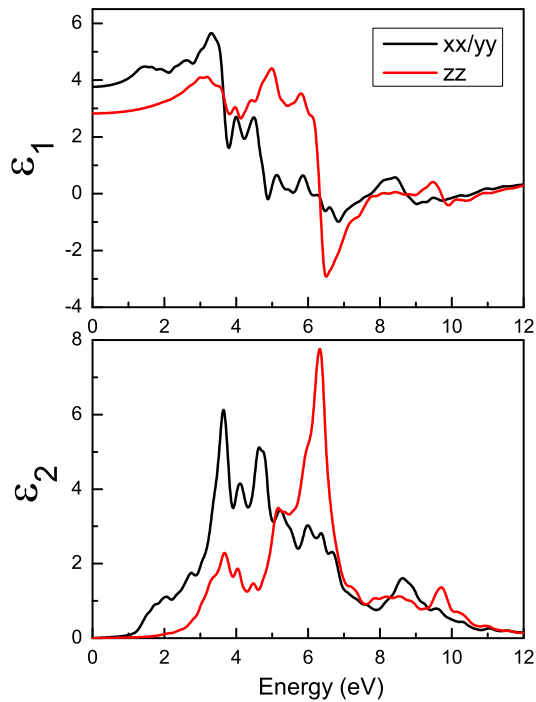}
  \caption{(Color online)Using GGA+SOC, calculated optical dielectric function of monolayer SnSSe, including real parts (Top) and imaginary
parts (Bottom) along xx/yy and zz directions. }\label{t511}
\end{figure}

\section{Optical properties}
The optical properties of the SnSSe monolayer can be described by the complex dielectric function $\varepsilon(\omega)$, the imaginary part $\varepsilon_2(\omega)$ of which   is determined by a summation
over empty band states as follows\cite{op1}:
\begin{equation}\label{op}
   \varepsilon_2(\omega)=\frac{2\pi e^2}{\Omega \epsilon_0}\sum_{k,v,c}\delta(E^c_k-E^v_k-\hbar\omega)|<\psi^c_k|u.r|\psi^v_k>|^2
\end{equation}
Where $\epsilon_0$, $\Omega$ and $\hbar\omega$ are the vacuum dielectric constant,  the volume and the energy of the incident phonon;
The $v$,  $c$ and $u$ represents the valence bands, conduction bands  and  the polarization vector in the incident electric field;
 The $u.r$ and  $\psi_k$ is the momentum operator and  the wave function at the k point. The real part  $\varepsilon_1(\omega)$ of
dielectric tensor can be  obtained by the Kramers-Kronig relation. According to the dielectric function, the  absorption coefficient $\alpha(\omega)$ can be given using the equation\cite{op2}:
\begin{equation}\label{op}
   \alpha(\omega)=\frac{\sqrt{2}\omega}{c}\{[\varepsilon_1^2(\omega)+\varepsilon_2^2(\omega)]^{1/2}-\varepsilon_1(\omega)\}^{\frac{1}{2}}
\end{equation}

\begin{figure}
    \includegraphics[width=7cm]{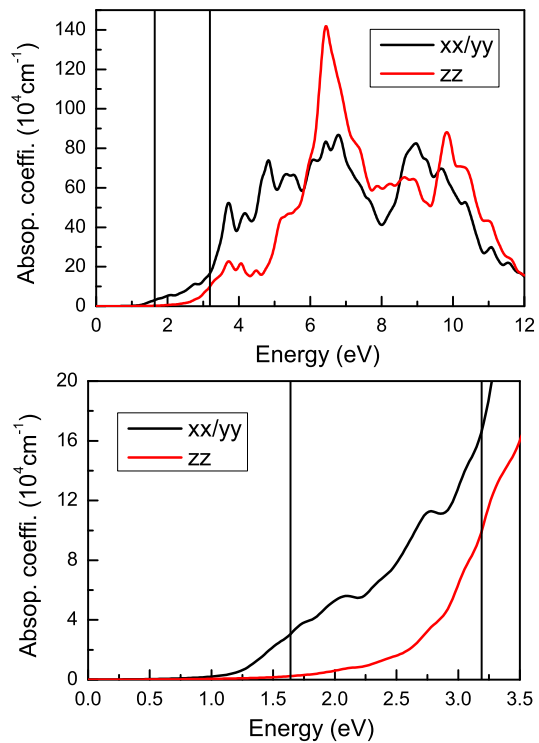}
  \caption{(Color online)Using GGA+SOC, the optical absorption coefficients of SnSSe monolayer along xx/yy and zz directions from 0 eV to 12 eV (Top) and from 0 eV to 3.5 eV (Bottom), and the visible light region (1.6-3.1 eV) is shown.  }\label{t512}
\end{figure}

Here, we use the rhombus primitive cell to investigate the optical properties of monolayer SnSSe. It is noted that the unit-cell volume $\Omega$ is not well-defined for 2D materials, and their optical properties depend on the the length of unit cell along z direction\cite{pv7}.
The dielectric function and absorption coefficient of SnSSe monolayer are calculated along xx/yy and zz directions, and are plotted in \autoref{t511} and \autoref{t512}.
Along xx/yy and zz directions, strong anisotropy in the optical spectra can be observed due to distinct optical selection rules.
In low frequency region, the  $\varepsilon_1(\omega)$ along xx/yy direction is higher than one along zz direction,
The  $\varepsilon_2(\omega)$  along xx/yy direction shows  a sharp increase at about 1.18 eV at the first onset of the optical transitions, which is
lower than one (2.00 eV) along zz directin. The main peak region of  $\varepsilon_2(\omega)$ along xx/yy direction centres  around 4 eV, and 6.3 eV along zz direction. From the ultraviolet to the visible light region, strong absorption intensity can be observed along xx/yy direction, and  the absorption coefficient values reach up to more than $10^4 cm^{-1}$ in the visible light region, implying high efficiency in the utilization
of solar energy. The absorption intensity along zz direction is lower than one along xx/yy direction in the visible light region. However, the very strong absorption intensity along zz direction is observed around 6.5 eV, which is higher than one along xx/yy direction.

\section{Electronic transport properties}
\begin{figure}
    \includegraphics[width=7cm]{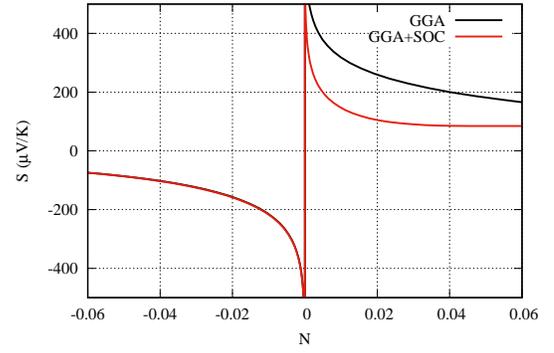}
  \caption{(Color online) The room-temperature transport coefficient S of SnSSe monolayer  as a function of doping level (N) using GGA and GGA+SOC.}\label{t5}
\end{figure}

\begin{figure}
    \includegraphics[width=8cm]{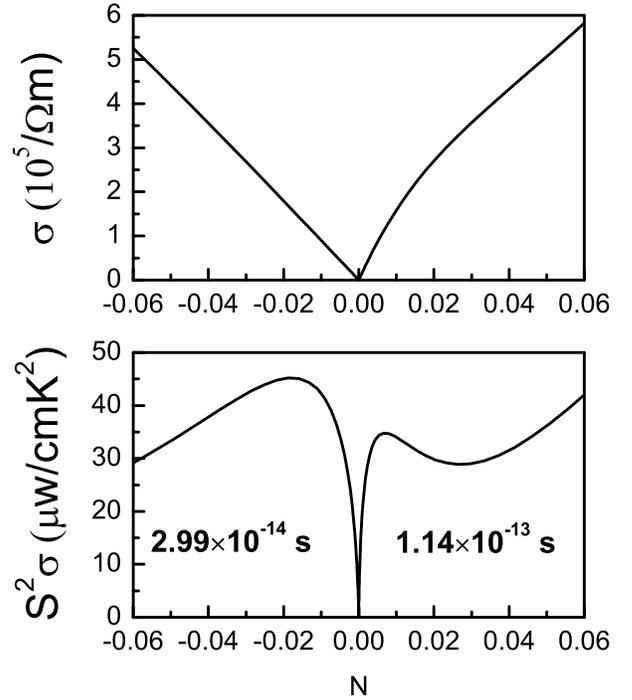}
  \caption{The room-temperature transport coefficients   $\mathrm{\sigma}$  and  $\mathrm{S^2\sigma}$ of SnSSe monolayer  as a function of doping level (N) using GGA+SOC.  }\label{t5-1}
\end{figure}

Based on CSTA Boltzmann theory within rigid band approach (RBA), the electronic  transport coefficients of SnSSe monolayer are calculated.
To simulate the doping effects, the n(p)-type doping is achieved by simply moving  the position of Fermi level into conduction  (valence) bands within RBA.  Here, the rhombus primitive cell is used to investigate the electronic transport properties of SnSSe monolayer, and the studied  transport direction is corresponding to zigzag (y) direction  of the rectangle supercell.
The  Seebeck coefficient S is independent of relaxation time  $\tau$, and the room temperature S as  a function of doping level (N) is shown in \autoref{t5} with both GGA and GGA+SOC.
The  SOC  can reduce  p-type  S of SnSSe monolayer, and has little  effect on S in n-type doping, which  can be understood   by considering SOC effects on the bands around the Fermi level. One reason is that the SOC can remove the valence  band degeneracy   near the VBM,  leading to reduced p-type S.  Another reason is that the SOC can reduce the effective masses of VBM (from \autoref{tab3}), producing a detrimental S in p-type doping due to:
\begin{equation}\label{s}
    S=\frac{8\pi^2K_B^2}{3eh^2}m^*T(\frac{\pi}{3n})^{2/3}
\end{equation}
where   $m^*$, T and  $n$ is  the effective mass of
the carrier, temperature and  carrier concentration, respectively. Thus, it is very important for electronic transport studies to consider SOC.
It is noted that the calculated electrical conductivity $\mathrm{\sigma/\tau}$ and  power factor $\mathrm{S^2\sigma/\tau}$ depend  on $\tau$.  By using calculated $\tau$ from \autoref{tab3}, the room temperature  $\mathrm{\sigma}$ and  $\mathrm{S^2\sigma}$  as  a function of N  are shown in \autoref{t5-1}, using  GGA+SOC. For n-type doping, the $\tau$=$2.99\times10^{-14}$ s, and $\tau$=$1.14\times10^{-13}$  s in p-type doping.  Calculated results show that SnSSe monolayer  has very excellent $\mathrm{S^2\sigma}$. It is found that the n-type doping has better $\mathrm{S^2\sigma}$ than p-type one.
Ultra low lattice thermal conductivity in monolayer $\mathrm{SnS_2}$ or $\mathrm{SnSe_2}$ has been predicted from a first principles study\cite{la1}.
The lattice thermal conductivity of Janus MoSSe/PtSSe/ZrSSe monolayer is lower than $\mathrm{MoS_2}$/$\mathrm{PtS_2}$/$\mathrm{ZrS_2}$ monolayer\cite{p4,p4-1,p4-2}. Thus, the SnSSe monolayer should have ultra low lattice thermal conductivity, which suggests that Janus monolayer SnSSe may be a potential thermoelectric
material.

\section{Discussions and Conclusion}

Monolayer $\mathrm{SnX_2}$ (X = S, Se) have been recently synthesized\cite{q6-1,q6-2}, and Janus TMD monolayer MoSSe with sandwiched S-Mo-Se structure has also been synthesized by replacing the top S atomic layer in  $\mathrm{MoS_2}$ with Se atoms\cite{p1,p1-new}. Thus, it is possible to achieve Janus SnSSe monolayer by replacing the top S (Se) atomic layer in  $\mathrm{SnS_2}$ ($\mathrm{SnSe_2}$) with Se (S) atoms. In fact, growing  Janus SnSTe and SnSeTe monolayers is also possible. Compared to $\mathrm{SnS_2}$ and $\mathrm{SnSe_2}$ monolayers, the most important difference is that  Janus SnXY (X/Y=S, Se, Te) monolayer breaks  inversion symmetry, and therefore piezoelectricity can be produced. In addition to this, the electronic structures, phonons, elastic properties, optical properties and transport properties of $\mathrm{SnS_2}$ and $\mathrm{SnSe_2}$ monolayers can be tuned by constructing Janus monolayer. It has been proved that $\mathrm{SnS_2}$ and $\mathrm{SnSe_2}$ monolayers have ultra low lattice thermal conductivities, due to the low phonon
velocity, low Debye temperature, weak bonding interactions  and strong anharmonicity\cite{la1}. The Janus SnXY (X/Y=S, Se, Te)  monolayer should also have ultra low lattice thermal conductivity, and they may be promising candidates for thermoelectric
applications.

 In summary, the electronic structures, carrier mobilities, piezoelectric properties, optical properties and  transport coefficients   of Janus SnSSe monolayer are systematically studied from  the reliable first-principle calculations. Janus monolayer SnSSe  with  1T phase is found to exhibit mechanical  and dynamic stability and high experimental feasibility. Calculated results show that SnSSe monolayer is an indirect gap semiconductor, and the SOC can produce important influences on energy band structures, effective masses of VBM  and electronic transport coefficients of Janus SnSSe monolayer. The  hole mobilities of monolayer SnSSe
are higher than  electron ones, and are also higher  than those of monolayer $\mathrm{SnS_2}$   or  $\mathrm{SnSe_2}$\cite{la1}.
The piezoelectric coefficient $d_{11}$  of SnSSe monolayer is comparable  with one of $\alpha$-quartz  commonly used
3D piezoelectric material\cite{yd3}. The high absorption coefficients in the visible light region can be observed along xx/yy direction.
The electronic transport calculations  suggest that Janus monolayer SnSSe may be a potential thermoelectric
material. Our works can stimulate further experimental works to synthesize  SnSSe  monolayer, and  will motivate farther   studies of  other Janus monolayers, such as SnSTe and SnSeTe monolayers.

\begin{acknowledgments}
This work is supported by the National Natural Science Foundation of China (Grant No. 11404391). We are grateful to the Advanced Analysis and Computation Center of China University of Mining and Technology (CUMT) for the award of CPU hours and WIEN2k/VASP software to accomplish this work.
\end{acknowledgments}

\end{document}